\documentclass[12pt]{article}
\usepackage{amsmath, amssymb, slashed, epsf, color}
\usepackage{geometry}
\usepackage{fullpage}
\usepackage{mathabx}
\usepackage{cancel}

\usepackage{graphicx}
\usepackage[colorlinks=true]{hyperref}

\newcommand{\beq}{\begin{equation}}
\newcommand{\eeq}{\end{equation}}

\begin{document}

%%%%%%%%%%%%%%%%%%%%%%%%%%%%%%%%%%
%%%%%%%%%%% Title page %%%%%%%%%%%
%%%%%%%%%%%%%%%%%%%%%%%%%%%%%%%%%%
\begin{titlepage}
\begin{center}
\hfill CPHT-RR022-0519 \\
\hfill UMN--TH--3823/19, FTPI--MINN--19/14 \\ 
\hfill  LPT-Orsay-19-23

\vspace{1.0cm}
{\Large\bf Limits on $R$-parity Violation in High Scale Supersymmetry}

\vspace{1.0cm}
Emilian Dudas$^{a}$,
Tony Gherghetta$^{b}$,
Kunio Kaneta$^{b,c}$,\\
Yann Mambrini$^{d}$,
and
Keith A. Olive$^{b,c}$

\vspace{0.5cm}
{\it
${}^a$CPHT, CNRS, Ecole Polytechnique, IP Paris, F-91128 Palaiseau, France \\
${}^b$School of Physics and Astronomy, University of Minnesota, \\Minneapolis, Minnesota 55455, USA\\
${}^c$William I.~Fine Theoretical Physics Institute, University of Minnesota, \\Minneapolis, Minnesota 55455, USA\\
${}^d$Laboratoire de Physique Th\'eorique (UMR8627), CNRS, Univ. Paris-Sud, \\Universit\'e Paris-Saclay, 91405 Orsay, France
}

\vspace{0.5cm}
\abstract{
We revisit the limits on $R$-parity violation in the minimal supersymmetric standard model. 
In particular, we focus on the high-scale supersymmetry scenario in which all the sparticles are in excess of the inflationary scale of approximately $10^{13}$ GeV, and thus no sparticles ever come into thermal equilibrium. 
In this case the cosmological limits, stemming from the preservation of the baryon asymmetry that 
have been previously applied for weak scale supersymmetry, are now relaxed.
We argue that even when sparticles are never in equilibrium, $R$-parity violation is still constrained via higher dimensional operators by neutrino and nucleon experiments and/or insisting on the preservation of a non-zero $B-L$ asymmetry.
}

\end{center}
\end{titlepage}
\setcounter{footnote}{0}

\section{Introduction}
\label{sec:introduction}

Operators which violate baryon and/or lepton number represent a two-edged sword for beyond the Standard Model
physics. On the one hand, some degree of baryon or lepton number violation is necessary in order to account
for the observed baryon asymmetry of the Universe. As is well known, these $C$ and $CP$-violating interactions
must be out-of-equilibrium to generate a non-zero asymmetry. An out-of-equilibrium decay, for example,
can generate a baryon or lepton asymmetry if the $C$, $CP$, and $B$ and/or $L$-violating decay occurs at a temperature
significantly below the mass of the decaying particle \cite{sw,ttwz}. A simple rule of thumb condition on the mass, $M$, of the decaying particle is $M > y^2 M_P$, where $y$ is the coupling leading to the decay and $M_P$ is the 
(reduced) Planck mass, $M_P^2 = 1/(8\pi G_N)$. In the case that the decay is purely lepton number violating, as in 
leptogenesis \cite{fy}, the lepton asymmetry must be converted at least in part to a baryon asymmetry
with sphaleron processes \cite{KRS,AMc}. However, sphaleron mediated interactions violate $B+L$ while conserving $B-L$ and
hence require a non-zero $B-L$ asymmetry to be generated. Most importantly, since the $B-L$ conserving sphaleron
processes remain in equilibrium up to the time of the electroweak phase transition,
any other process in equilibrium which violates another combination of $B$ and $L$ would lead to the complete
wash-out of any baryon or lepton asymmetry independent of its origin. This allows one to place strong 
constraints on any possible $B$ and/or $L$-violating operators \cite{ht,fy2,nb,cdeo1,cdeo2,fglp,iq}.

These constraints are particularly important in supersymmetric models with $R$-parity violation (RPV)\cite{Farrar:1978xj}.
Indeed, $R$-parity is usually imposed in supersymmetric models to avoid fast baryon and lepton number violating
interactions which could lead to rapid proton decay. Limits on proton stability are satisfied if new interactions 
violated either $B$ {\em or} $L$. However, if these interactions remain in equilibrium
at the same time as sphalerons are in equilibrium, they would wash away any baryon asymmetry \cite{cdeo1,cdeo2}
despite proton stability.
One should bear in mind, that these bounds can be evaded if there is a residual lepton number conservation \cite{dr,cdeo3},
or if the lepton asymmetry is stored in a Standard Model (SM) $SU(2)$ singlet such as the right-handed electron \cite{cdeo3,cko},
or other flavor asymmetries \cite{krs2,dr,dko}.
However, the wash out can be affected by slepton mixing angles~\cite{endo}.

Unfortunately, the absence of a supersymmetry signal at the LHC \cite{nosusy}, means that the scale of supersymmetry 
remains unknown. While naturalness can be used to argue for supersymmetry at or near the weak scale \cite{Maiani:1979cx},
the supersymmetry breaking scale may turn out to be much larger. 
The order parameter for supersymmetry breaking is related to the gravitino mass, $m_{3/2}$, 
and in  minimal anomaly-mediated supersymmetry breaking models \cite{anom,ggw,mAMSB},
the gravitino mass is typically several hundred TeV to $\mathcal{O}(1)$ PeV. 
In these models, gaugino masses are loop suppressed with respect to the gravitino mass,
and scalar masses may be considerably lighter. 
In models of split supersymmetry \cite{split} and in models of pure gravity mediation \cite{pgm}, the gravitino mass
and scalar masses may lie beyond the PeV scale.
In models of high-scale supersymmetry \cite{hssusy,hssusy2,egko}, the gravitino and sparticle masses may be even higher.
When the gravitino mass is $\mathcal{O}(1)$ EeV, a new window opens up for gravitino dark matter
when all of the sparticle masses (except the gravitino mass) lie above the inflationary scale of 
approximately $10^{13}$ GeV \cite{bcdm,eev,LH}. However, without some degree of RPV,
it is hard to imagine experimental tests to detect EeV gravitino dark matter when all sparticle masses
are $> 10^{13}$ GeV.

In high-scale supersymmetry, the limits on RPV are relaxed as the supersymmetric particles
were never in the thermal bath and could not participate in interactions that wash out the baryon asymmetry. Therefore, some amount of RPV is acceptable, and if present, RPV operators would render the lightest supersymmetric particle, the gravitino in this case, unstable.
If long-lived, the decay products may provide a signature for an EeV gravitino \cite{LH}.
A smoking gun signal could occur from EeV monochromatic neutrinos observable by
IceCube and/or ANITA \cite{IC, ANITA}.
This, in fact is a generic prediction, because given the milder assumptions on RPV couplings, high-scale supersymmetric scenarios are more naturally $R$-parity violating, allowing more general UV completions where $R$-parity conservation is not necessary. This compares with weak scale supersymmetry where RPV couplings need to be significantly suppressed, that either requires an additional suppression mechanism or strongly suggests that $R$-parity is conserved. 

In deriving the limits on RPV parameters we distinguish between two cases depending on whether or not the 
gravitino is the dark matter.  The limits on RPV parameters are generally stronger when the 
gravitino lifetime is required to be long enough so as to allow for gravitino dark matter. When these 
limits are not satisfied, an alternative to gravitino dark matter is required in high-scale supersymmetric models. 

In this note we discuss the cosmological limits on the RPV interactions in high-scale supersymmetry models where all 
sparticles are assumed to have never been in chemical equilibrium with the SM particles.
Thus we are able to derive new constraints on RPV operators in models of high-scale supersymmetry. 
The outline of the paper is as follows: In section \ref{sec:cosmological_limits}, we review previous 
cosmological constraints on generic higher-dimension operators that arise from the preservation of the baryon asymmetry. 
In section \ref{sec:laboratory_limits} we review and update experimental limits arising from neutrino masses, nucleon decay and $n-\bar{n}$ oscillations. Specific RPV interactions are discussed in section \ref{sec:r_parity_violating_interactions}, and we derive new limits for high-scale supersymmetry which are then  compared with those from weak scale supersymmetry. Finally
we also discuss limits that arise from including the gravitino. A summary of our results is given in section \ref{sec:summary}.

\section{Cosmological Limits} 
\label{sec:cosmological_limits}

Our constraints on RPV interactions are derived from the requirement that $B-L$ violating interactions are not in equilibrium simultaneously with sphaleron processes when they are operative.
The sphaleron rate is estimated at next-to-leading order in, refs. \cite{Bodeker:1999gx,Arnold:1999uy,Buchmuller:2005eh}, 
and sphaleron processes are in thermal equilibrium at temperatures 
\begin{eqnarray}
	T_c \lesssim T \lesssim T_{\rm sph},
	\label{eq: Tc<T<Tsph}
\end{eqnarray}
where $T_{\rm sph}\simeq 10^{12}$ GeV and $T_c\simeq 160$ GeV is the critical temperature of the electroweak phase transition~\cite{DOnofrio:2015gop}.
Thus, if there is a $B-L$ violating process in thermal equilibrium at some time when the temperature $T$ satisfies (\ref{eq: Tc<T<Tsph}), any baryon/lepton number asymmetry is washed out unless it is regenerated after the electroweak phase transition. 
As noted earlier, there are exceptions to this very general criterion \cite{dr,cdeo3,cko,krs2,dko}.
Nevertheless, we are interested in deriving general bounds and 
to preserve a non-zero asymmetry, we will require all $B-L$ violating processes to be decoupled
in the range given by (\ref{eq: Tc<T<Tsph}). In practice we can distinguish two cases depending on the 
temperature dependence of the $B/L$ violating rate, $\Gamma_{BL}$.  Assuming that $\Gamma_{BL} \propto \,T^{2D-7}$(see below),  we require the $B/L$ violating rate to be less than 
the Hubble expansion rate, $\Gamma_{BL} < H$ at $ T = T_c$ for $D \leq 4$ (corresponding to relevant or marginal operators) [case (a)], while for $D >4$ (corresponding to irrelevant operators), it is necessary for  $\Gamma_{BL} < H$ at $ T =T_{\rm sph}$  [case (b)]. 

\subsection{Thermal equilibrium constraints}
Here, we focus on high-scale supersymmetry scenarios in which all superpartners with the exception 
of the gravitino have masses greater than the inflationary scale (inflaton mass). As a result
these particles were not produced during reheating and were never in thermal equilibrium.
To describe particle interactions in a thermal bath we can use effective operators consisting of only SM particles~\cite{cdeo1}. These can be written as 
\begin{eqnarray}
	{\cal L}_D = \frac{{\cal O}_D}{M_D^{D-4}} \, ,
\end{eqnarray}
where $M_D$ represents an effective heavy particle mass scale, and
${\cal O}_D$ is an operator with mass dimension $D$.
Then, the reaction rate of such an interaction is given by
\begin{eqnarray}
	\Gamma_D \sim c_D\left(\frac{T}{M_D}\right)^{2(D-4)} T \, ,
\end{eqnarray}
with a prefactor $c_D$ (to be explained shortly). This rate should be compared with the Hubble scale $H=0.33 g_*^{1/2}T^2/M_{P} \equiv T^2/\widetilde M_P$.
As there are only SM particles in the thermal bath, the number of relativistic degrees of freedom, $g_* = 427/4$,  and $\widetilde M_P\simeq 7.04\times10^{17}$ GeV.
Thus our limits on $M_D$ (which contain all couplings in addition to the heavy mass scale) become
\begin{eqnarray}
M_D & > & \left(c_D{\widetilde M_P} T_{\rm c}^{2D-9} \right)^{\frac{1}{2(D-4)}} \qquad D < 4\, , \label{Mlimit1} \\
M_D & > &  \left(c_D{\widetilde M_P} T_{\rm sph}^{2D-9} \right)^{\frac{1}{2(D-4)}} \qquad D > 4 \, .
\label{Mlimit2}
\end{eqnarray}

Next, we consider the reaction rate $\Gamma_D$, which can be expressed more accurately as
\begin{eqnarray}
	\Gamma_D &=& \frac{N^2 T}{32\pi^4n_0}\int_0^\infty ds\, s^{3/2} 
	\sigma(2\to k)K_1(\sqrt{s}/T),
\end{eqnarray}
where $\sigma(2\to k)$ is the $2\to k$ scattering cross section, $K_1$ is a modified Bessel function, $n_0=N \zeta(3)T^3/\pi^2$ is the initial particle number density with $\zeta(3) \simeq 1.202$ and $N$ the number of degrees of freedom for the initial state particle. Note we have neglected the difference between Fermi and Bose statistics, and all of the initial and final state particles are assumed to be massless.
The cross sections are proportional to $s^{D-5}$ for the operators with $D\geq4$, and thus we obtain
\begin{eqnarray}
	\int_0^\infty ds\, s^{3/2}s^{D-5} K_1(\sqrt{s}/T) &=& 2^{2(D-3)}T^{2D-5}\Gamma(D-3)\Gamma(D-2)\, ,
\end{eqnarray}
where $\Gamma$ is the Gamma function.
A concrete form of $\sigma(2\to k)$ depends on, for example, the spinor and derivative structures of corresponding operators.
However, since listing a complete set of $B-L$ breaking operators is not the aim of this paper, we only keep track of a typical phase space volume factor in the following argument.
Thus, we may write
\begin{eqnarray}
	\sigma(2\to k) &\simeq& \frac{1}{2s}|{\cal M}(2\to k)|^2 \Phi_k \, ,
\end{eqnarray}
where $\Phi_k$ is the $k$-body phase space volume which is approximately given as
\begin{eqnarray}
	\Phi_k &\simeq& \frac{1}{8\pi}\left(\frac{s}{16\pi^2}\right)^{k-2} \, ,
\end{eqnarray}
with $k\geq 2$.\footnote{We implicitly assume the $s$-channel type decomposition for multi-particle final state diagrams. For more details, see, e.g., Ref.~\cite{Kersevan:2004yh}.}

For the $D=3,4$ operators, one-to-two processes are also possible.
When we parametrize the decay width for a particle with mass $M$ to two Dirac fermions by
\begin{eqnarray}
	\Gamma(1\to2) &\equiv& \frac{\lambda^2}{8\pi}M \, ,
\end{eqnarray}
with a generic coupling $\lambda$, the reaction rate for $D=3,4$ becomes
\begin{eqnarray}
	\Gamma_D &\simeq& \frac{N}{2\pi^2n_0}MT^2\Gamma(1\to2)
	=\frac{\lambda^2}{16\zeta(3)\pi}\frac{M^2}{T},
	\label{decay}
\end{eqnarray}
and thus we have $c_3, c_4(\text{decay})=\lambda^2/(16\zeta(3)\pi)$.
For a two-to-two process with quartic coupling $\lambda$,
we find
\begin{eqnarray}
\Gamma_D &\simeq&
	\frac{\lambda^2 N}{128\zeta(3)\pi^3}T,
	\label{2to2}
\end{eqnarray}
where $c_4$(scattering) = $\lambda^2 N/(128\zeta(3)\pi^3)$ for two-to-two processes\footnote{This result differs from that in \cite{cdeo2} by a factor of $6/\pi^2~ (2) $ for the decay (two-to-two) processes that results from our approximation of using Maxwell-Boltzmann statistics in the thermal average.}.

Table \ref{tab:reaction_rate} summarizes the decay width and cross sections along with the coefficients $c_D$ which are used to limit $M_D$ for each 
operator, and where $\psi, \phi, {\cal D}$ symbolically denote a fermion, scalar, and covariant derivative, respectively.
Again, in the table we omit prefactors in the cross sections coming from the kinematics and coupling structure.
For dimension seven operators involving a covariant derivative, we consider only derivative terms, since they are dominant compared to the terms with a gauge boson.
The resultant limits on the operators are summarized in Table \ref{tab:limits} where $q,l (l^c), h$ symbolically represent quark, lepton (charge conjugate of a lepton field), and the SM Higgs fields, respectively.

\begin{table}[th!]
\begin{center}
	\begin{tabular}{cccc}
		$D$ & Operator & $\Gamma(1\to2)$& $c_D$\\\hline
		3 & $\phi^3$ & $\frac{\lambda^2}{8\pi}M$ & $\frac{\lambda^2}{16\zeta(3)\pi}$\\
		4 & $\phi\psi^2$ & $\frac{\lambda^2}{8\pi}M$ & $\frac{\lambda^2}{16\zeta(3)\pi}$\\\\
		$D$ & Operator & $\sigma(2\to k)$& $c_D$\\\hline
		4  & $\phi^4$ & $ \frac{1}{16\pi}\frac{\lambda^2}{s}$ \;\;\; $(k=2)$ & $\frac{\lambda^2}{128\zeta(3)\pi^3}$\\
		5 & $\psi^2 \phi^2$ & $\frac{1}{16\pi}\frac{1}{M_5^2}$ \;\;\; $(k=2)$ & $\frac{1}{16\zeta(3)\pi^3}$\\
		6 & $\psi^4$ & $\frac{1}{16\pi}\frac{s}{M_6^4}$ \;\;\; $(k=2)$ & $\frac{3}{2\zeta(3)\pi^3}$\\
		7 & $\psi^4 \phi$ & $\frac{1}{16\pi(4\pi)^2}\frac{s^2}{M_7^6}$  \;\;\; $(k=3)$ & $\frac{9}{2\zeta(3)\pi^5}$\\
		  & $\psi^4{\cal D}$ & $\frac{1}{16\pi}\frac{s^2}{M_7^6}$  \;\;\; $(k=2)$ & $\frac{72}{\zeta(3)\pi^3}$\\
		9 & $\psi^6$ & $\frac{1}{16\pi(4\pi)^4}\frac{s^4}{M_9^{10}}$  \;\;\; $(k=4)$ & $\frac{2700}{\zeta(3)\pi^7}$ \\
		10 & $\psi^6 \phi$ & $\frac{1}{16\pi(4\pi)^6}\frac{s^5}{M_{10}^{12}}$  \;\;\; $(k=5)$ & $\frac{28350}{\zeta(3)\pi^9}$ 
	\end{tabular}
	\caption{Cross sections and prefactors in the reaction rates are shown for each operator with effective mass scale $M_D$. Although we do not consider dimension six operators when applying cosmological limits, as they do not violate $B-L$, we show them in the Table for completeness. For $D=3,4$, we denote $\lambda$ as a generic coupling constant and $M$ as the mass of the decaying particle. We have neglected factors of $N$ related to the number of degrees of freedom of incoming states and used Maxwell-Boltzmann statistics in the thermal average.}
	\label{tab:reaction_rate}
\end{center}
\end{table}

\begin{table}[ht!]
\begin{center}
	\begin{tabular}{ccccccc}
		$D$ & Operator & $\Delta B$ & $\Delta L$ & \multicolumn{2}{c}{Cosmological limits}  & Laboratory limits \\ \cline{5-6}
		& & & & case (a) & case (b) & \\ \hline
		5 & $llhh$ & 0 & 2 & $M_5 > 10^{8.6}$ GeV & $M_5 > 10^{13.5}$ GeV & $M_5>10^{14.7}$ GeV\\
		6 & $qqql$ & 1 & 1 & none & none & $M_6>10^{15.8}$ GeV\\
		7 & $\psi^4 h (qqql^ch, {\rm etc.})$ & \multicolumn{2}{c}{$\cancel{B-L}$} & $M_7 > 10^{4.5}$ GeV & $M_7 > 10^{12.6}$ GeV & $M_7>10^{11.1}$ GeV \\
		& $\psi^4 {\cal D} (qqql^c{\cal D}, {\rm etc.})$ & \multicolumn{2}{c}{$\cancel{B-L}$} & $M_7 > 10^{4.8}$ GeV & $M_7 > 10^{13.0}$ GeV & $M_7>10^{10.1}$ GeV \\
		9 & $qqqqqq$ & 2 & 0 & $M_9 > 10^{3.8}$ GeV & $M_9 > 10^{12.6}$ GeV & $M_9>10^{5.9}$ GeV \\
		& $qqqlll$ & 1 & 3 & $M_9 > 10^{3.8}$ GeV & $M_9 > 10^{12.6}$ GeV & $M_9>10^{5.6}$ GeV\\
		10 & $qqql^cl^cl^ch$ & 1 & $-3$ & $M_{10} > 10^{3.5}$ GeV & $M_{10} > 10^{12.5}$ GeV & $M_{10}>10^{5.1}$ GeV
	\end{tabular}
	\caption{Cosmological and laboratory limits on baryon and/or lepton number violating operators.
	The cosmological limits assume the SM value of $g_* = 427/4$ as appropriate for high-scale supersymmetry.
	}
	\label{tab:limits}
\end{center}
\end{table}

\section{Laboratory Limits}
\label{sec:laboratory_limits}

In addition to the cosmological limits discussed above, there are of course a variety of 
laboratory limits on baryon and lepton number violating operators which we summarize in this section.
We will concentrate on limits from neutrino masses (on lepton number violating operators) and
nucleon decay limits. 

\subsection{Neutrino mass constraints}

We can derive a lower bound on the mass scale $M_5$ used in
the dimension five (Weinberg) operator, $(\bar L^c\cdot H)(L\cdot H)/M_5$, in addition to the cosmological bound in Eq.\,(\ref{Mlimit2}),  
from the upper bound on the sum over the neutrino masses, $\sum_i m_{\nu_i} < 0.15$ eV (95\% CL) \cite{sumnu}. 
In addition, we can derive an 
upper bound on $M_5$ from neutrino oscillation data.
To be more concrete, in the following discussion we assume that the heaviest neutrino 
mass is given by $m_\nu =v^2/M_5$ where $v=1/\sqrt{2\sqrt{2}G_F}\simeq 174$ GeV is the vacuum expectation value (VEV) of the SM Higgs field\footnote{Note that the Yukawa coupling, $y$, has been absorbed into the definition of $M_5$, so that if the Weinberg operator arises from a right-handed neutrino mass and the see-saw mechanism then $M_5 = M_R/y^2$.}.

Since the sign of the squared neutrino mass difference cannot be determined from atmospheric neutrino observations, the neutrino mass ordering can either be the normal hierarchy (NH) or the inverted hierarchy (IH).
The NH spectrum is defined by the neutrino mass ordering given by $m_{\nu_3}>m_{\nu_2}>m_{\nu_1}$, whereas the IH spectrum is
given by $m_{\nu_2}>m_{\nu_1}>m_{\nu_3}$.
The central values for the experimentally determined mixing angles and squared mass differences are \cite{Tanabashi:2018oca}\footnote{For $\sin^2 \theta_{23}$
we are using fits in the 2nd octant.},
\begin{eqnarray}
	&&\sin^2\theta_{12} = 0.307,~~~\sin^2\theta_{23} = 0.542~({\rm NH}),~~~0.536~({\rm IH}),~~~\sin^2\theta_{13} = 2.18\times10^{-2},\nonumber\\
	&&m_S^2 \equiv m_{\nu_2}^2 - m_{\nu_1}^2 = 7.53\times10^{-5}~{\rm eV}^2, \nonumber\\
	&&m_A^2 \equiv |m_{\nu_3}^2 - m_{\nu_2}^2| = 2.53\times10^{-3}~{\rm eV}^2~({\rm NH}),~~~2.44\times10^{-3}~{\rm eV}^2~({\rm IH}).
\end{eqnarray}
For the neutrino masses, there is only one free parameter in each case, namely, the lightest neutrino mass which we denote $m_{\nu 0}$.
Then, each mass spectrum is given by
\begin{eqnarray}
	{\rm NH}: && m_{\nu_1}=m_{\nu 0},~~~m_{\nu_2} = \sqrt{m_{\nu 0}^2 + m_S^2},~~~m_{\nu_3} = \sqrt{m_A^2+m_S^2+m_{\nu 0}^2},\\
	{\rm IH}: && m_{\nu_1}=\sqrt{m_A^2-m_S^2+m_{\nu 0}^2},~m_{\nu_2}=\sqrt{m_A^2+m_{\nu 0}^2},~m_{\nu_3}=m_{\nu 0}.
\end{eqnarray}

The limit $\sum_i m_{\nu_i} < 0.15$ eV sets an upper bound, $m_{\nu 0}\lesssim 4.16\times10^{-2}$ eV ($3.24\times10^{-2}$ eV) corresponding to the heaviest neutrino mass, $m_{\nu_3 (\nu_2)}\simeq 6.59\times10^{-2}$ eV ($5.91\times10^{-2}$ eV) for the NH (IH) case.
In both cases we obtain, $M_5\gtrsim10^{14.7}$ GeV. This constraint is listed in the first line of Table \ref{tab:limits}. 

On the other hand,  neutrino oscillations imply non-zero neutrino masses which gives an upper bound to $M_5$,
assuming that the only source for generating neutrino masses is the dimension five operator.
To obtain a conservative limit, we take $m_{\nu 0}=0$ in both cases, and then the heaviest mass becomes $m_{\nu_3 (\nu_2)}\simeq 5.10\times10^{-2}$ eV ($4.94\times10^{-2}$ eV) for the NH (IH) case. It turns out that in both cases we need $M_5\lesssim 10^{14.8}$ GeV to explain the neutrino oscillation data, provided the dimension five operator is the dominant contribution to the neutrino masses.\footnote{If there is another source for the neutrino masses and the dimension five operator is not dominant, then this upper limit does not apply, and $M_5$ can be as large as possible, e.g., $M_P$.}
The combined upper and lower limits point to a unique value in the range of $10^{14.7}~{\rm GeV}\lesssim M_5 \lesssim 10^{14.8}$ GeV, though one
needs to bear in mind that $M_5$ is an effective mass parameter for the heaviest neutrino that includes all relevant couplings.

\subsection{Nucleon decays}

Clearly, most baryon number violating operators can be constrained by proton or neutron decay.
Since the dimension six operator $qqql$ does not violate $B-L$, there is no cosmological limit from the preservation of the baryon asymmetry.
On the other hand, as baryon number is violated, the nucleon lifetime can be used to derive a lower bound on $M_6$.

In general, there are four types of such operators \cite{Weinberg:1979sa,Wilczek:1979hc,Abbott:1980zj}\footnote{Our definition of these operators is identical to that of Ref.\cite{Nagata:2013sba} after arranging $SU(2)$ and spin indices appropriately.}:
\begin{eqnarray}
	{\cal O}^{(1)}_{ijkl} &=& G^{(1)}_{ijkl}(\bar d^c_{i\alpha} P_R u_{j\beta})(\bar Q^c_{k\gamma}\cdot L_l)\epsilon^{\alpha\beta\gamma},\\
	{\cal O}^{(2)}_{ijkl} &=& G^{(2)}_{ijkl}(\bar Q^c_{i\alpha}\cdot Q_{j\beta})(\bar u^c_{k\gamma} P_R e_l)\epsilon^{\alpha\beta\gamma},\\
	{\cal O}^{(3)}_{ijkl} &=& G^{(3)}_{ijkl}(\bar Q^c_{i\alpha}\cdot Q_{j\beta})(\bar Q^c_{k\gamma} \cdot L_l)\epsilon^{\alpha\beta\gamma},\\
	{\cal O}^{(4)}_{ijkl} &=& G^{(4)}_{ijkl}(\bar d^c_{i\alpha} P_R u_{j\beta})(\bar u^c_{k\gamma} P_R e_l)\epsilon^{\alpha\beta\gamma},
\end{eqnarray}
where $i,j,k,l$ and $\alpha,\beta,\gamma$ are flavor and color indices respectively, and the $SU(2)$ product is denoted as $A\cdot B =\epsilon^{ab}{A_a}_L{B_b}_L$ ($\epsilon^{12}=\epsilon^{123}=1$), i.e., $\bar Q^c\cdot L = \bar u^c P_L e-\bar d^c P_L \nu$.
All fermions are defined as four component spinors, and the $SU(2)$ doublets include the chiral projection $P_L$ as appropriate.
The flavor dependent Wilson coefficients are represented by $G^{(n)}_{ijkl}$ ($n=1,2,3,4$), whose flavor structure depends on the underlying theory.
Instead of specifying a concrete flavor structure, we assume that there are neither any degenerate parameters (i.e., no accidental cancellations) nor large hierarchies among the different flavor entries.

For proton decay involving a charged antilepton in the final state, the strongest limit arises from the decay mode $p\to e^+\pi^0$, which is given by $\tau_{p\to e\pi}>1.6\times10^{34}$ years \cite{Miura:2016krn}.
The relevant operators for this decay channel are
\begin{eqnarray}
	&&G^{(1)}_{1111}(\bar d^c P_R u)(\bar u^c P_L e),~~~
	-2G^{(2)}_{1111}(\bar d^c P_L u)(\bar u^c P_R e),\nonumber\\
	&&-2G^{(3)}_{1111}(\bar d^c P_L u)(\bar u^c P_L e),~~~
	G^{(4)}_{1111}(\bar d^c P_R u)(\bar u^c P_R e),
\end{eqnarray}
where the flavor mixing from the CKM matrix is neglected.
The decay width of this process is given by
\begin{eqnarray}
	\Gamma(p\to e^+\pi^0) &=& \frac{m_p}{32\pi}
	\left(
		1-\frac{m_\pi^2}{m_p^2}
	\right)^2
	\left(
		|{\cal A}_L|^2 + |{\cal A}_R|^2
	\right)\, ,
\end{eqnarray}
where $m_p (m_\pi)$ is the proton (pion) mass. The amplitudes are defined as\footnote{Note that in the specific case of SU(5), proton decay mediated by $X$ and $Y$ gauge bosons only
involves $G^{(1)}$ and $G^{(2)}$ \cite{hkn}.}
\begin{eqnarray}
	{\cal A}_L &=& G^{(1)}_{1111}\langle\pi^0|(ud)_Ru_L|p\rangle - 2G^{(3)}_{1111}\langle\pi^0|(ud)_L u_L|p\rangle,\\
	{\cal A}_R &=& -2G^{(2)}_{1111}\langle\pi^0|(ud)_Lu_R|p\rangle + G^{(4)}_{1111}\langle\pi^0|(ud)_R u_R|p\rangle,
\end{eqnarray}
where we have used the notation
\begin{eqnarray}
	(ud)_\Gamma u_{\Gamma'} &=& \epsilon^{\alpha\beta\gamma}(u^T_\alpha CP_\Gamma d_\beta)P_{\Gamma'}u_\gamma\,,
\end{eqnarray}
with $\Gamma, \Gamma'=L,R$, and $\alpha, \beta, \gamma$ are $SU(3)$ color indices \cite{Aoki:2013yxa}.
Then, by taking $G^{(1,2,3,4)}_{1111}\sim 1/M_6^2$ and the hadron matrix elements $\langle\pi^0|(ud)_{L,R}u_{L,R}|p\rangle 
\sim -\langle\pi^0|(ud)_{L,R} u_{R,L}|p\rangle 
\sim 0.1~{\rm GeV}^2$ \cite{Aoki:2013yxa}, we obtain $M_6 > 10^{15.8}$ GeV. This constraint is listed in the second line of 
Table \ref{tab:limits}. 
If some of the dimension six operators are absent, this constraint is somewhat relaxed.
For instance, when ${\cal O}^{(1)}$ is the sole operator mediating proton decay, the limit reduces to $M_6 > 10^{15.5}$ GeV.

A similar procedure can be applied for the operators whose mass dimension is greater than six.
\begin{enumerate}
\item {\bf Dimension seven operators:} The $qqql^ch$ type operators may be regarded as a dimension six operator where the Wilson coefficient, estimated as $\sim v/M_7^3$, leads to the limit $M_7\gtrsim 10^{11.1}$ GeV.
In the same way, the $qqql^c{\cal D}$ type operators may be regarded as dimension six operators with the Wilson coefficient $\sim \Lambda_{\rm QCD}/M_7^3$ with $\Lambda_{QCD}\simeq 300$ MeV \cite{Tanabashi:2018oca}, and thus we obtain $M_7\gtrsim 10^{10.1}$ GeV.
It should be noted that for dimension seven operators, there are additional operators that violate $B-L$, such as $(\bar e L\cdot H)(\bar L^c \cdot L)$.\footnote{There also exist operators involving more than two Higgs doublets or covariant derivatives (equivalently, gauge field strength), which we do not discuss here.} 
It is also known that all dimension seven operators that violate either baryon or lepton number, violate $B-L$ as well.
For more on all such operators, see Refs. \cite{Lehman:2015coa,Henning:2015alf}.

\item {\bf Dimension nine operators:} The $qqqlll$ operator may also induce nucleon decays.
For instance, the operator $(\bar L^c\cdot L)(\bar e P_L d^c)(\bar Q\cdot Q^c)$ causes $n\to e^+e^-\nu$ which has the lifetime constraint $\tau_{n\to e^+e^-\nu}>2.8\times10^{32}$ years \cite{McGrew:1999nd}.
Then, we obtain $M_9>10^{5.6}$ GeV where we have estimated the decay width as $\Gamma(n\to e^+e^-\nu)\simeq \alpha_h^2m_n^{5}/(256\pi^3M_9^{10})$ (by only taking into account the phase space volume) with the hadron matrix element $\alpha_h\simeq-0.0144~{\rm GeV}^3$ \cite{Aoki:2017puj}.\footnote{This constraint was missing in Ref. \cite{Costa:1985vk}, and an accurate estimate has recently been given in Ref. \cite{Faroughy:2014tfa} which also argues that $n\to K^0l^+l^-\nu$, induced by, for example, $(\bar L^c\cdot L)(\bar e P_L d)(\bar u P_L d^c)$, is larger than $n\to l^+l^-\nu$ in RPV models, although this particular channel is not yet constrained by experiments (see also Ref. \cite{Hambye:2017qix}).}
The $qqqqqq$ type of operators, especially $uddudd$, are constrained by $n-\bar n$ oscillation.
Following Refs. \cite{Mohapatra:1980qe,Mohapatra:1989ze}, the $n-\bar n$ mixing time can be written as $\tau_{n-\bar n}\simeq 1/\delta m$ with
\begin{eqnarray}
	\delta m \sim \frac{1}{M_9^5} |\psi(0)|^2\, ,
\end{eqnarray}
where $\psi(0)$ denotes the neutron wave function at the origin, which is typically $\psi(0)\sim\Lambda_{\rm QCD}^3$.
The current constraint $\tau_{n-\bar n}>2.7\times 10^8$ s \cite{Abe:2011ky} sets the limit $M_9>10^{5.9}$ GeV.

\item {\bf Dimension ten operators:} For instance, $(\bar L^c\cdot L)(\bar e P_L d)(\bar u \,Q^c\cdot H)$ induces the nucleon decay $n\to e^+e^-\nu$ whose decay width may be evaluated as $\Gamma_{n\to e^+e^-\nu}\simeq \alpha_h^2m_n^{5}v^2/(256\pi^3 M_{10}^{12})$, and thus we obtain $M_{10}>10^{5.1}$ GeV.
\end{enumerate}

Finally let us reiterate that for operators of mass dimension higher than seven there exist many baryon and/or lepton number violating operators which are not listed in Table \ref{tab:limits}.
Though the $D=7, 9, 10$ operators in Table \ref{tab:limits} are just examples, it is sufficient for our purpose since such higher dimensional operators would usually involve more undetermined parameters compared to lower dimensional ones, and thus the detailed constraints are strongly model dependent.
For instance, some operators that violate lepton number, but conserve baryon number, can be constrained by neutrinoless double beta decay \cite{deGouvea:2007qla}, while the limit strongly depends on the form of the operators.
Nevertheless, when we assume that baryon and lepton numbers are violated at the same scale, and that there is no large hierarchy between the mass scales of baryon and lepton number violation, the constraints on $M_D$ from nucleon decays are in most cases stronger than those from neutrinoless double beta decay (see, e.g., Ref. \cite{Herrero-Garcia:2019czj}.)

\section{$R$-parity Violating Interactions} 
\label{sec:r_parity_violating_interactions}

We now discuss the limits on RPV interactions using the results obtained in the previous sections.
The RPV superpotential is given by
\begin{eqnarray}
	W_{\rm RPV} &=& W_{\rm RPV}^{(2)} + W_{\rm RPV}^{(3)},\\
	W_{\rm RPV}^{(2)} &=& \mu'_i H_u\cdot L_i, \label{bilinear}\\
	W_{\rm RPV}^{(3)} &=& \frac{1}{2}\lambda_{ijk}L_i\cdot L_j E^c_k + \lambda'_{ijk} L_i \cdot Q_i D^c_k + \frac{1}{2}\lambda''_{ijk}U^c_i D^c_j D^c_k\,. \label{quartic}
\end{eqnarray}
The explicit Lagrangian including soft supersymmetry breaking terms is shown in  the Appendix.
We will first review the bounds derived in the case of weak scale supersymmetry \cite{cdeo2} and contrast them with bounds obtained in high-scale supersymmetry. These bounds are derived from both the cosmological preservation of the 
baryon asymmetry and the experimental limits on baryon and/or lepton number violating processes including proton decay.
We will also comment on the limits on the RPV parameters when we require a sufficiently long-lived gravitino 
as the dark matter. 

In general the RPV mass parameter $\mu'_i$ depends on lepton flavor, but here we omit the flavor dependence for simplicity, and take $\mu'_i\equiv \mu'$.
(For a more detailed discussion, see, e.g., \cite{Barbier:2004ez,Diaz:1997xc}.)
Since lepton number is not conserved, $L$ and $H_d$ cannot be distinguished, and thus there is a field basis dependence in defining $L$ and $H_d$ fields.
For instance, if $L \to (1 - \epsilon^2)^{1/2} L + \epsilon H_d$ and $H_d \to (1 - \epsilon^2)^{1/2} H_d - \epsilon L$ 
with $\epsilon=\mu'/\sqrt{\mu^2 + \mu'^2}$ and $\mu$ is the $\mu$-parameter in the MSSM superpotential, we can eliminate the bilinear RPV term at the expense of generating trilinear RPV terms, such as $y_u \epsilon LLE^c$ and $y_d \epsilon QLD^c$. For simplicity and since observables do not depend on the choice of basis, we will work in the basis that explicitly keeps 
the bilinear term (\ref{bilinear}) given in $W_{\rm RPV}$.

\subsection{Limits on $\mu'$}
\subsubsection{Weak scale supersymmetry}
As discussed above, there are strong constraints on baryon and lepton number violating 
operators whose
induced interactions are simultaneously in equilibrium with the sphaleron interactions.
In the case of an $R$-parity violating bilinear $L H_u$ term, one-to-two processes involving a Higgsino, lepton, and a gauge boson will be induced.  From Eq.\,(\ref{decay}), the thermally averaged rate at a temperature, $T$ for these lepton number violating interactions is given by \cite{cdeo2,LH}
\beq
\Gamma_{1\rightarrow 2} = \frac{g^2 \theta^2 T}{16 \zeta(3) \pi} \simeq 
0.016 g^2 \frac{\mu'^2}{m_f^2}T \, ,
\label{12rate}
\eeq
where $g$ is a gauge coupling, and $\theta \simeq \mu'/m_f$ is the mixing
angle induced by $\mu'$
for a fermion with mass $m_f$. We require that this lepton number violating interaction is out of equilibrium. As such, we require the interaction rate (\ref{12rate}) is less than the Hubble rate, $H \simeq  \sqrt{\pi^2 g_*/90}~ T^2/M_P$.
This implies that
\beq
\mu'^2 < 20 \sqrt{g_*} \frac{T^3}{M_P} \, ,
\label{m'limit}
\eeq
where the fermions have a thermal mass, $m_f \sim g T$.
We further insist that any lepton number violating rate involving $\mu'$ remains out of equilibrium while sphaleron interactions are in equilibrium, i.e., between the weak scale $T_c$ and $T_{sph}$.  As one can see, the limit (\ref{m'limit}) is strongest for $T$ of order the weak scale (case (a) corresponding to $D=3$).  For weak scale supersymmetry, the fermion can be either a lepton or Higgsino,  $g_* = 915/4$
and at $T_c$ one obtains the limit \cite{cdeo2}
\beq
\mu' < 2.3 \times 10^{-5} \,{\rm GeV} \, .
\label{Eq:limit2}
\eeq
For 
weak scale supersymmetry this limit translates to $\epsilon \lesssim 2.3 \times 10^{-7}$.

In general, the RPV bilinear term induces a non-zero neutrino mass via a dimension five operator.
The mixing angle between neutrinos and the Higgsino is given by $\mu'/\mu$, and 
through the Higgsino-Higgs-gaugino (wino or bino) coupling, we obtain a dimension five operator of the form:
\begin{eqnarray}
	{\cal L}_5 \simeq \frac{1}{M_5} \nu_L\nu_L h h,~~~
	\frac{1}{M_5} \simeq \epsilon^2
	\frac{g_2^2 M_1 + g_1^2 M_2}{M_1M_2(1+\tan^2\beta)}\,,
	\label{eq:RPV_dim5}
\end{eqnarray}
where $M_1 (M_2)$ are the bino (wino) masses and $g_2 (g_1)$ is the $SU(2)_L (U(1)_Y)$ gauge coupling.

In weak scale supersymmetry models, the limit (\ref{Eq:limit2}) is stronger than the limit
from neutrino masses  \cite{Barbier:2004ez,isy} which comes from the dimension five operator with the constraint given in Table \ref{tab:limits}.
 As one can see from Table \ref{tab:limits}, the strongest limit from a dimension five two-to-two process is obtained by
requiring the out-of-equilibrium condition to hold at the highest possible scale, which in this case is $T_{sph}$ (case (b)). 
For weak scale supersymmetry, the limit on $M_5$ becomes
\beq
M_5 > \frac{(c_5 T M_P)^{1/2}}{\sqrt{0.33 g_*^{1/2}}} \approx 2.8 \times 10^{13}\, {\rm GeV} \, ,
\eeq
for $T = T_{sph}$ and $g_* = 915/4$ (the change in $g_*$ accounts for the slight difference with respect to the limit in Table~\ref{tab:limits}).
This translates to the limit 
\beq
\mu' < 1.9 \times 10^{-7} {\rm GeV}^{-1/2} {\widetilde m}^{1/2} \mu (1 + \tan^2 \beta)^{1/2}/g \approx  4.4 \times 10^{-4}\, {\rm GeV} \, , 
\eeq
for $\mu \sim M_1 \sim M_2 \sim {\widetilde m} \sim 100$ GeV, and $\tan \beta \approx 1$. 
We assume a generic gauge coupling $g\sim 0.6$ throughout. In this case, $\epsilon \lesssim 4.4 \times 10^{-6}$.

\subsubsection{High-scale supersymmetry}
\label{sec:bilinearHighScale}
In the case of high-scale supersymmetry,  we assume that all  sparticles are heavier than the inflationary mass scale $m_I \sim 3 \times 10^{13}$ GeV, 
and we denote the typical sparticle mass scale as $\widetilde m > m_I$.
As all sparticle masses are greater than $T_{sph}$, there are no sparticles in the thermal bath when sphalerons are in equilibrium and the limit from one-to-two processes is not applicable.
Nevertheless, the limit from the effective dimension five operator is valid when the heavy sparticles are integrated out.
Since only Standard Model particles are in the thermal bath, $g_* = 427/4$ and we can use the limit on $M_5$ from
Table \ref{tab:limits} (case (b)). The limit on $\mu'$ becomes
\beq
\mu' < 1.7 \times 10^{-7} {\rm GeV}^{-1/2} {\widetilde m}^{1/2} \mu (1 + \tan^2 \beta)^{1/2}/g \approx 6.6 \times 10^{13} \,{\rm GeV} \, , 
\eeq
for $\mu \sim M_1 \sim M_2 \sim {\widetilde m} \sim 3 \times 10^{13}$ GeV, and $\tan \beta \approx 1$. In this case, $\epsilon \lesssim 2.2$.

As one can also see from Table {\ref{tab:limits}, the laboratory limit in this case is in fact the strongest limit on $\mu'$.
Using $M_5 > 5 \times 10^{14}$ GeV, we obtain
\beq
\mu' < 4.5 \times 10^{-8} {\rm GeV}^{-1/2} {\widetilde m}^{1/2} \mu (1 + \tan^2 \beta)^{1/2}/g \approx 1.7 \times 10^{13}\, {\rm GeV} \, , 
\eeq
or $\epsilon \lesssim 0.57$.

Note that if $W^{(2)}_{\rm RPV}$ is the only source of neutrino mass, our previous limit on $M_5 < 10^{14.8}$ GeV translates into a lower bound on $\mu'$,
\beq
\mu' >  4 \times 10^{-8} {\rm GeV}^{1/2} {\widetilde m}^{1/2} \mu (1 + \tan^2 \beta)^{1/2}/g \approx 1.5 \times 10^{13} \,{\rm GeV} \, .
\eeq
As discussed in Section \ref{sec:laboratory_limits}, the lower limit can be removed if there is another source for generating neutrino masses that can explain the neutrino oscillation data.

\subsection{Limits on $\lambda,\lambda',\lambda''$}
\subsubsection{$D=4,5$}
The quartic couplings in Eq. (\ref{quartic}) can lead to either one-to-two processes (involving a scalar and two fermions) or two-to-two processes
(involving four scalars) which violate baryon and/or lepton number. 
The rates for these processes taken from Table \ref{tab:reaction_rate} can be written as \cite{cdeo2}
\begin{eqnarray}
\Gamma_{2\rightarrow 2} & = & \frac{\lambda^2 y^2 T}{128 \zeta(3) \pi^3} \simeq 
2 \times 10^{-4} \lambda^2 y^2 T \, , \label{quartic12} \\
\Gamma_{1\rightarrow 2} & = &  \frac{\lambda^2 m_0^2 }{16 \zeta(3) \pi T} \simeq 
0.016 \lambda^2 \frac{m_0^2}{T} \label{quartic22} \, ,
\label{1222rate}
\end{eqnarray}
where $\lambda$ is a generic RPV quartic coupling in (\ref{quartic}) and $m_0 < T$ is the 
scalar mass. The rate (\ref{quartic12}) depends on the Standard Model Yukawa coupling $y$, because the baryon/lepton number violating processes actually arise from a cross term in the 
$F$-term in the scalar potential. 

In weak scale supersymmetry, these processes will be in equilibrium unless $\lambda$ is quite small, and the limit on $\lambda$ is 
derived by comparing these rates with the Hubble rate. 
This yields the limits
\begin{eqnarray}
\lambda  & <  & 1.2 \times 10^{-6} y^{-1} \qquad 2 \leftrightarrow 2\, , \label{qlimit} \\
\lambda & <  & 1.4 \times 10^{-7} \qquad\quad 1 \leftrightarrow 2 \, ,
\end{eqnarray}
where we have evaluated the limit at $T \sim m_0 \sim T_c$ in the one-to-two rate. 

Once again, in the case of high-scale supersymmetry, when all sparticle masses are greater than the inflationary scale,
the above limits are no longer applicable as there are no sparticles in the thermal bath at the time 
when sphaleron interactions are in equilibrium.
For the RPV bilinear term, we were able to derive a limit on $\mu'$ by integrating out the heavy sparticles
and setting a limit on the resulting dimension five operator. One might think that one can do the same for the quartic coupling,
and form a dimension six (four-fermion) operator and still set (weaker) limits on the RPV quartic couplings. However, as shown in  
\cite{Weinberg:1979sa}, there are no $B-L$ violating dimension-six operators involving only Standard Model fields.

There are, however, numerous laboratory and astrophysical constraints on the RPV quartic couplings which are independent of the sphaleron processes
\cite{qlab,sd,bm,DH,bs,bgh,Zwirner:1984is}. 
For example, some of the quartic couplings will contribute radiatively to neutrino masses and neutrinoless double 
beta decay \cite{qlab}, where these limits scale as $\lambda < \mathcal{O}(10^{-3}) ({\widetilde m}/100 {\rm GeV})^p$ for $p = 1/2, 5/2$ respectively.
As one can see, in the high-scale supersymmetric limit, the bounds on these couplings also disappear. 
The same is true of collider limits \cite{DH} and cosmological and astrophysical limits from the decay of the lightest supersymmetric particle \cite{bs}.

Furthermore let us comment on the issue of radiatively induced neutrino masses \cite{radmnu}.
Possible radiative corrections through the RPV couplings are summarized in Ref.~\cite{Davidson:2000ne}, where the relevant contribution in our case is the self-energy diagrams (diagram 19 in that paper) involving $\mu'$ and $B'(\equiv B_i)$.
The correction to the neutrino mass, $\delta m_\nu$, is proportional to $\mu' B'/\widetilde m^2$.
However, once all the Higgs boson contributions are incorporated, 
one finds that $\delta m_\nu$ is suppressed by $(v/\widetilde m)^2$, and thus $\delta m_\nu$ may be written in terms of the dimension five operator with a loop factor, i.e., $(16\pi^2)^{-1}(\bar L^c \cdot H)(L\cdot H)/M_5$ with $1/M_5\sim \mu' B'/\widetilde m^4\sim\mu'^2/\widetilde m^3$.
Therefore, the constraint from $\delta m_\nu$ is weaker than that coming from the tree level (dimension five) operator.

\subsubsection{$D=6$}
Despite the weakening of most bounds on the RPV quartic couplings, 
there remain limits on dimension six operators which induce proton decay.\footnote{See \cite{RPVPD} for related discussion on proton decay constraints.}
Once again, since these operators conserve $B-L$, there are no limits from the sphaleron wash-out of the baryon asymmetry.
Nevertheless,  proton decay is induced by $\tilde d^c$ exchange diagrams in the RPV case \cite{Weinberg:1979sa}, 
and only ${\cal O}^{(1)}$ type of operators can appear.
The corresponding Wilson coefficients are
\begin{eqnarray}
	G^{(1)}_{ijkl} \simeq \sum_{m,n=1}^3\lambda''^*_{jim} \lambda'_{lkn} (m^{-2}_{\tilde d^c})_{mn},
\end{eqnarray}
where the relation $\lambda''_{ijk}=-\lambda''_{ikj}$ is imposed by gauge symmetry.
By ignoring flavor mixing in the down-type squark sector, we obtain 
\begin{eqnarray}
	{\cal O}^{(1)}_{1111} = \frac{1}{M_6^2} (\bar d^c_1 P_R u_1)(\bar Q^c_1\cdot L_1),~~~
	\frac{1}{M_6^2} \sim \frac{1}{\widetilde m^2}\sum_{m=1}^3\lambda''^*_{11m}\lambda'_{11m},
\end{eqnarray}
and thus the limit on $M_6$ from Table \ref{tab:limits} can be expressed as the following limit 
on the quartic coupling
\begin{eqnarray}
	& & \left|\sum_{m=1}^3\lambda''^*_{11m}\lambda'_{11m}\right| < 2.3\times10^{-5} \left(\frac{\widetilde m}{3 \times 10^{13}~{\rm GeV}}\right)^2\, ,
\end{eqnarray}
which updates the results given in Ref. \cite{Barbier:2004ez}.

\subsubsection{$D=7$}
Dimension seven operators of the type, $qqql^ch$, 
are induced by involving trilinear couplings, i.e., $A^d_{ij} h_d\cdot \tilde Q_i \tilde d^c_j$ and $\mu y^{d*}_{ij}h_u^\dagger\cdot \tilde Q_i \tilde d^c_j$.
Then, we have
\begin{eqnarray}
	{\cal L}_7^{(1)} &=& G_{7,ijkl}^{(1)}(\bar L_i\cdot h_d d_j)(\bar Q^c_l\cdot Q_k),\\
	{\cal L}_7^{(2)} &=& G_{7,ijkl}^{(2)}(\bar L_i\cdot h_u^\dagger d_j)(\bar Q^c_l\cdot Q_k),
\end{eqnarray}
with coefficients
\begin{eqnarray}
	G_{7,ijkl}^{(1)} &\simeq& \lambda'^*_{imj}(m^{-2}_{\tilde Q})_{mm'}A^d_{m'n'}(m^{-2}_{\tilde d^c})_{n'n}\lambda''^*_{kln},\\
	G_{7,ijkl}^{(2)} &\simeq& \lambda'^*_{imj}(m^{-2}_{\tilde Q})_{mm'}\mu^* y^d_{m'n'}(m^{-2}_{\tilde d^c})_{n'n}\lambda''^*_{kln},
\end{eqnarray}
respectively, where $\mu$ is assumed to be complex.
These operators give rise to interaction rates which scale as 
\beq
\Gamma_7 \simeq c_7 (\lambda'\lambda'')^2 \left(\frac{A^2}{{\widetilde m}^8}\right) T^7 \, ,
\eeq
with $c_7=9/2\zeta(3)\pi^5$ given in Table \ref{tab:reaction_rate}, and $A$ denotes an $A$-term.
When compared to the Hubble rate, one sees that the appropriate limit, evaluated at $T = T_{sph}$ (case (b)), gives
\begin{eqnarray}
	& & \left|\sum_{m=1}^3 \lambda'^*_{imj}\lambda''^*_{klm}\right| < 290 \left(\frac{\widetilde m}{3 \times 10^{13}~{\rm GeV}}\right)^4\left(\frac{3 \times 10^{13}~{\rm GeV}}{|A^d_0|}\right),\\
	& & \left|\sum_{m,n=1}^3\lambda'^*_{imj}y^d_{mn}\lambda'^*_{klm}\right| <290 \left(\frac{\widetilde m}{3 \times 10^{13}~{\rm GeV}}\right)^4\left(\frac{3 \times 10^{13}~{\rm GeV}}{|\mu|}\right),
\end{eqnarray}
where $(m_{\tilde Q}^2)_{ij}\sim(m_{\tilde d^c}^2)_{ij}\sim \widetilde m^2\delta_{ij}$ and $A^d_{ij} = A^d_0\delta_{ij}$
and we have assumed that there is no flavor mixing in the soft supersymmetry breaking terms.
These cosmological limits are stronger than the nucleon decay limits which are 
\begin{eqnarray}
	& & \left|\sum_{m=1}^3 \lambda'^*_{1m1}\lambda''^*_{11m}\right| < 10^7\left(\frac{\widetilde m}{3\times 10^{13}~{\rm GeV}}\right)^4\left(\frac{3 \times 10^{13}~{\rm GeV}}{|A^d_0|}\right),\\
	& & \left|\sum_{m,n=1}^3\lambda'^*_{1m1}y^d_{mn}\lambda'^*_{11m}\right| < 10^7\left(\frac{\widetilde m}{3 \times 10^{13}~{\rm GeV}}\right)^4\left(\frac{3 \times 10^{13}~{\rm GeV}}{|\mu|}\right),
\end{eqnarray}
and become very weak in the high-scale supersymmetric limit.

Through the trilinear couplings $A^u_{ij}h_u\cdot\tilde Q_i\tilde u^c_j$ and $\mu y^{u*}_{ij}h_d^\dagger\cdot\tilde Q_i\tilde u^c_j$, we also have
\begin{eqnarray}
	{\cal L}_7^{(3)} &=& G_{7,ijkl}^{(3)}(\bar L_i\cdot h_u d_j)(\bar d^c_l\cdot d_k),\\
	{\cal L}_7^{(4)} &=& G_{7,ijkl}^{(4)}(\bar L_i\cdot h_d^\dagger d_j)(\bar d^c_l\cdot d_k),
\end{eqnarray}
with coefficients
\begin{eqnarray}
	G_{7,ijkl}^{(3)} &\simeq& \lambda'^*_{imj}(m^{-2}_{\tilde Q})_{mm'}A^u_{m'n'}(m^{-2}_{\tilde u^c})_{n'n}\lambda''^*_{kln},\\
	G_{7,ijkl}^{(4)} &\simeq& \lambda'^*_{imj}(m^{-2}_{\tilde Q})_{mm'}\mu^* y^u_{m'n'}(m^{-2}_{\tilde u^c})_{n'n}\lambda''^*_{kln},
\end{eqnarray}
and the constraints on $G_{7,ijkl}^{(3)}$ and $G_{7,ijkl}^{(4)}$ can be obtained in the same way.

\subsubsection{$D=9,10$}
RPV interactions also induce $n-\bar n$ oscillations via dimension nine operators, 
which can be written in the following form \cite{Zwirner:1984is},
\begin{eqnarray}
	{\cal L}_9 &\supset& G_{9,ijklmn} (\bar d^c_iP_R u_j)(\bar d^c_k P_R u_l)(\bar d^c_m P_R d_n)\,.
\end{eqnarray}
There are two possible diagrams that produce this operator, namely, via the $A$-term or gluino exchange.
In each case, we obtain
\begin{eqnarray}
	G_{9,ijklmn}(A-\rm{term}) 
	&\simeq& 
	\sum_{ss'tt'uu'}\frac{\lambda''^*_{jis}\lambda''^*_{lkt}\lambda''^*_{unm}A''_{s't'u'}}{(m^2_{\tilde u^c})_{ss'}(m^2_{\tilde d^c})_{tt'}(m^2_{\tilde d^c})_{uu'}},\\
	G_{9,ijklmn}(\text{gluino})
	&\simeq&
	\sum_{m'n'}\frac{g_s}{M_3}\frac{\lambda''^*_{ijm'}\lambda''^*_{lkn'}}{(m^2_{\tilde d^c})_{m'm}(m^2_{\tilde d^c})_{n'n}}\,,
\end{eqnarray}
where $M_3$ is the gluino mass, $g_s$ is the QCD coupling and $A''$ is a soft mass parameter (see Appendix). The rate for these processes can be approximated as
\begin{eqnarray}
\Gamma_9 & = & c_9\lambda''^6 \left(\frac{A''^2}{{\widetilde m}^{12}} \right) T^{11}\,, \qquad A-{\rm term} \\
\Gamma_9 & = & c_9\lambda''^4 g_s^2 \left(\frac{1}{{\widetilde m}^{10}} \right) T^{11}\,, \qquad {\rm gluino} 
\end{eqnarray}
with $c_9=2700/\zeta(3)\pi^7$ given in Table \ref{tab:reaction_rate}.
Then, once again comparing to the Hubble rate at $T = T_{sph}$, we obtain the constraints as follows:
\begin{eqnarray}
	& &
	\left|\sum_{stu} \lambda''^*_{jis}\lambda''^*_{lkt}\lambda''^*_{unm}\left(\frac{A''_{stu}}{3\times10^{13}~{\rm GeV}}\right)\right| < 3.4\times10^{4}\left(\frac{\widetilde m}{3\times10^{13}~{\rm GeV}}\right)^6,\\
	& &
	\left|\lambda''^*_{ijm}\lambda''^*_{lkn}\right| < 5.6\times10^{4}\left(\frac{\widetilde m}{3\times10^{13}~{\rm GeV}}\right)^5,
\end{eqnarray}
where $M_3\sim\widetilde m$ and $(m^2_{\tilde u^c})_{ij}\simeq(m^2_{\tilde d^c})_{ij}\sim \widetilde m^2\delta_{ij}$.
It is also true that the above case (b) limit leads to a stronger bound than the $n-\bar n$ oscillation limits which are effectively absent,
\begin{eqnarray}
	& &
	\left|\sum_{stu} \lambda''^*_{11s}\lambda''^*_{11t}\lambda''^*_{u11}\left(\frac{A''_{stu}}{3\times10^{13}~{\rm GeV}}\right)\right| < 7.7 \times 10^{37}\left(\frac{\widetilde m}{3\times10^{13}~{\rm GeV}}\right)^6,\\
	& &
	\left|\lambda''^*_{111}\lambda''^*_{111}\right| < 1.3\times10^{38}\left(\frac{\widetilde m}{3\times10^{13}~{\rm GeV}}\right)^5.
\end{eqnarray}
We also have dimension nine operators of the type $qqqlll$, which can be expressed as
\begin{eqnarray}
	{\cal L}_9 &\supset& G_{9,ijklmn}(\bar L^c_i\cdot L_j)(\bar e_kP_L d^c_l)(\bar u_mP_L d^c_n),\\
	G_{9,ijklmn} &\simeq& \sum_{ss'tt'uu'}\lambda_{sjk}\lambda'_{itl}\lambda''_{mnu}(m_{\tilde L}^{-2})_{ss'}(m_{\tilde Q}^{-2})_{tt'}(m_{\tilde d^c}^{-2})_{uu'}A'^*_{s't'u'}.
\end{eqnarray}
The decay $n\to K l\bar l\nu$ occurs through this operator, although 
this particular channel has not been  experimentally constrained.
The decay $n\to l\bar l\nu$ happens through $(\bar L^c_i\cdot L_j)(\bar e_kP_L d^c_l)(\bar Q_m\cdot Q^c_n)$ in dimension nine.
However, as discussed in Ref. \cite{Faroughy:2014tfa}, such operators should be loop suppressed.
This may be understood by comparing with the operator $(\bar L^c_i\cdot L_j)(\bar e_kP_L d^c_l)(\bar u_mP_L d^c_n)$ in which one of the $d^c$ quarks should be an $s$ quark because of $SU(3)$ color symmetry. A chiral flip is needed to avoid the appearance of $s$ quark.
For instance, once the diagram involves a Higgs boson loop, $(\bar u_mP_L d^c_n)$ may be replaced by $y^u y^d/16\pi^2\times(\bar Q_m\cdot Q^c_n)$ which is less restrictive for the RPV couplings.

We may also construct a dimension ten operator of the type $qqql^cl^cl^ch$ by looking at dimension nine operators.
For example, $d^c_n$ in $(\bar L^c_i\cdot L_j)(\bar e_kP_L d^c_l)(\bar u_mP_L d^c_n)$ may be replaced by $m_d/v\times (Q^c_n\cdot H)$, to yield $(\bar L^c_i\cdot L_j)(\bar e_kP_L d^c_l)(\bar u_m Q^c_n\cdot H)$.
However, for these higher dimensional operators, the cosmological limit is much stronger than the laboratory limits as seen in the dimension seven operator case.

\subsection{Limits from the gravitino}

The limits on $\mu'$ in high-scale supersymmetry from Section~\ref{sec:bilinearHighScale} are based solely on the preservation of the baryon
asymmetry and experimental limits on the neutrino mass.  However, 
if we insist that the lightest supersymmetric particle, the gravitino, is relatively stable so that
it can play the role of dark matter, we can derive a significantly stronger limit on $\mu'$ \cite{LH}.
The presence of the RPV parameter $\mu'$ opens up channels for gravitino decay into
neutrinos plus gauge/Higgs bosons. 
The total decay rate is \cite{LH}
\begin{equation}
	\Gamma_{3/2} \simeq \frac{\epsilon^2 \cos^2 \beta m_{3/2}^3}{16\pi M_P^2},
\end{equation}
where $\epsilon \approx \mu'/\mu \approx \mu'/m_I$ and $\cos \beta \approx 1/\sqrt{2}$. 
Demanding that the gravitino lifetime exceeds the current age of the universe ($\tau_{3/2} > 4.3 \times 10^{17}$ s) 
corresponds to a limit on $\mu' < 0.03$ GeV, for $m_I = 3\times 10^{13}$ GeV and $m_{3/2} = 1$ EeV.
Note that the upper limit on $\mu'$ scales as $m_I/m_{3/2}^{3/2}$.
An even more restrictive limit on $\mu'$ is possible from the IceCube constraints on the neutrino flux produced by the gravitino decay \cite{IC}. In this case, we require a lifetime $\tau_{3/2} >  10^{28}$ s which corresponds to a limit, $\mu' < 2 \times 10^{-7}$ GeV. As one can see, these limits are far more restrictive than those from baryon/lepton number conservation, and neutrino masses.

While the quartic RPV couplings can also induce gravitino decay,
they do so only at the one-loop level.  As a consequence, the limit on a generic quartic coupling is very weak. For example, writing a generic dimension six operator as $(fff\psi_\mu)/M^2_6$ with $f$ and $\psi_\mu$ being SM fermions and the gravitino, respectively, the corresponding cosmological limit on $M_6$ from the preservation of $B-L$ asymmetry is
\begin{eqnarray}
	M_6 &>& (c_6 \widetilde M_P T_{sph}^3)^{1/4}\simeq 1.3\times10^{13}~{\rm GeV}\,,
	\label{climitm6}
\end{eqnarray}
with $c_6=3/2\zeta(3)\pi^3$ given in Table \ref{tab:reaction_rate}.
The same dimension six operator makes the gravitino unstable, and the gravitino decay width into three SM fermions is estimated as
\begin{eqnarray}
	\Gamma_{3/2 \to3f} &\simeq& \frac{m_{3/2}^5}{256\pi^3M^4_6}.
	\label{32decay}
\end{eqnarray}
Assuming the gravitino lifetime is longer than the age of the Universe gives rise to the lower bound
\begin{eqnarray}
	M_6 &>& 5.3\times10^{20}~{\rm GeV}\left(\frac{m_{3/2}}{\rm EeV}\right)^{5/4} \qquad (\tau_{3/2}>4.3\times 10^{17}~{\rm s})\,.
	\label{lifelimitm6}
\end{eqnarray}
Furthermore, the operator $(\bar d^c P_L u)(\bar d^c \gamma^\mu \psi_\mu)$ can be induced at the one-loop level, with a suppression scale $M_6^{-2}\sim |\lambda''|^2A''/(16\pi^2\tilde m^2 M_P)\sim |\lambda''|^2/(16\pi^2\tilde m M_P)$.
From the cosmological limit (\ref{climitm6}), this gives
\begin{eqnarray}
	|\lambda''| &\lesssim& 8.2\times10^{3},
\end{eqnarray}
when $\tilde m\sim m_I = 3\times10^{13}$ GeV.
Again imposing the gravitino lifetime limit (\ref{lifelimitm6}), we obtain
\begin{eqnarray}
	|\lambda''| &\lesssim& 2\times10^{-4}\left(\frac{\rm EeV}{m_{3/2}}\right)^{5/4}.
\end{eqnarray}
 In this case, the limit on the quartic coupling is only competitive with the limit in (\ref{qlimit}) when the gravitino mass begins to 
approach the inflationary scale, or when $m_{3/2} \gtrsim 0.002\times m_I$. 
If instead we impose the IceCube gravitino lifetime limit, $\tau_{3/2}>10^{28}$ s \cite{IC}, the generic constraint becomes
\begin{eqnarray}
	M_6 &>& 2.1 \times 10^{23}~{\rm GeV} \left(\frac{m_{3/2}}{\rm EeV}\right)^{5/4} \;\;\; (\tau_{3/2}>10^{28}~{\rm s}),
	\label{alimitm6}
\end{eqnarray}
and thus we obtain
\begin{eqnarray}
	|\lambda''| &\lesssim& 5\times10^{-7}\left(\frac{\rm EeV}{m_{3/2}}\right)^{5/4}.
\end{eqnarray}
This limit is more stringent than (\ref{qlimit}) when $m_{3/2} \gtrsim 1.7\times10^{-5}m_I$.

Finally note that there may be additional RPV operators which are non-renormalizable corrections to the
superpotential and the K\"ahler potenital. These operators can also contribute to the baryon number violating
interactions.  For example, consider the operator in Eq.\,(2.72) of \cite{Barbier:2004ez} with coupling $\kappa_7$.
By introducing a supersymmetry breaking spurion superfield $X$ (containing the Goldstino $\psi$), we may write the corresponding term as
\begin{eqnarray}
	K &\supset& \frac{\kappa_7}{M_P^2}Q\cdot Q d^{c\dagger}X^\dagger,
\end{eqnarray}
which gives $(\kappa_7/M_P^2)(\bar d^c Q)\cdot(\bar\psi Q)$.
Then by identifying $M_6=M_P/\sqrt{\kappa_7}$, the cosmological limit becomes 
$\kappa_7\lesssim 3.4 \times10^{10}$.
This will lead to a dimension nine operator induced by gravitino exchange, which remains below the Hubble rate so long as
$\kappa_7<2.1\times10^{11}$.\footnote{Here we assume $m_{3/2}<T_{sph}$ and estimate the reaction rate as $\kappa_7^4T^9/M_P^8$ which should be compared with $H$ at $T=T_{sph}$.}  However, this operator could
also lead to gravitino decay and using Eq.(\ref{32decay}), we can derive a much stronger limit, $\kappa_7<2 \times10^{-5}({\rm EeV}/m_{3/2})^{5/2}$ for $\tau_{3/2}>4.3\times10^{17}$ s, and $\kappa_7<1.4\times10^{-10}({\rm EeV}/m_{3/2})^{5/2}$ for $\tau_{3/2}>10^{28}$ s.
We do not consider these non-renormalizable operators any further.

\section{Summary}
\label{sec:summary}

We have revisited the limits on RPV in the minimal supersymmetric standard model, where the constraints from laboratory experiments and the preservation of the $B-L$ asymmetry are discussed.
In particular, we have focused on the high-scale supersymmetry scenario in which all the sparticles are in excess of the inflationary scale of approximately $10^{13}$ GeV, and thus they were never in equilibrium.
Since the previously argued cosmological limits in weak-scale supersymmetry assume that the sparticles involved in the RPV interactions remain in equilibrium at $T_c$, those limits cannot be applied to the high-scale supersymmetry case, and thus, the cosmological limits from the preservation of a non-zero $B-L$ asymmetry are relaxed.
Nevertheless, even when sparticles are never in equilibrium, RPV couplings are still constrained through higher dimensional operators.

Based on effective operators, we have reviewed and updated the experimental and cosmological limits on $B$ and/or $L$ violating processes, and then applied them for RPV in the high-scale supersymmetry scenario.
For dimension five operators, we have shown that the neutrino mass constraints are stronger than the cosmological limit, while for operators of mass dimension higher than seven, the cosmological limit is stronger than the experimental limits.
Dimension six operators are only constrained by nucleon decay experiments since there are no $B-L$ violating operators of dimension six.
We have contrasted the limits on RPV in high-scale supersymmetry with those in weak-scale supersymmetry up to dimension ten operators, and shown that indeed a wider range of RPV couplings is acceptable. This implies that unlike weak-scale supersymmetry, high-scale supersymmetry can generically have RPV with mild constraints on the couplings and imposing an $R$-parity is not necessary. This then leads to the generic prediction of an EeV gravitino decay. 

We have also distinguished limits based on the assumption of gravitino dark matter.
In this case, the RPV interactions lead to the possibility of gravitino decay.
If long-lived, the gravitino may still provide a sufficient mass density to make up the dark matter.
Indeed, if very long lived, present day decays may yet provide a signature.
For example, the RPV bilinear proportional to $\mu'$ induces a decay
to neutrinos which could be seen in high energy neutrino detectors \cite{LH}.
If we require the presence of dark matter today, or sufficiently long lived so as 
not to surpass the existing experimental constraints, we obtained limits on $\mu'$ 
which are far stronger than those from baryon/lepton number violation.
The RPV quartic couplings on the other hand are better constrained by baryon/lepton number 
violating rates.  We also noted that our limits can be applied to non-renormalizable corrections
in supergravity models, with the most stringent limits arising from gravitino decay.

%%%%%%%%%%%%%%%%%%%%%%%%%%%%%%%%%%%%%%%
%%%%%%%%%%% Acknowledgments %%%%%%%%%%%
%%%%%%%%%%%%%%%%%%%%%%%%%%%%%%%%%%%%%%%
\section*{Acknowledgments}
We would like to thank N. Nagata for helpful communications.
This  work was supported by the France-US PICS MicroDark.
 Y.M. acknowledges partial support from the European Union Horizon 2020
research and innovation programme under the Marie Sklodowska-Curie: RISE
InvisiblesPlus (grant agreement No 690575), the ITN Elusives (grant
agreement No 674896) and the Red Consolider  MultiDark 
FPA2017-90566-REDC. E.D. acknowledges partial support from the ANR
Black-dS-String. The work of T.G., K.K., and  K.A.O. was supported in
part by the DOE grant DE--SC0011842 at the University of Minnesota.

\appendix
\section*{Appendix: Notations and Lagrangian} 
\label{sec:notations_and_lagrangian}

We summarize the relevant Lagrangian we have used in our discussion for the sake of completeness.
Our notations and conventions basically follow Appendix A and B of Ref.\cite{Barbier:2004ez}.~\footnote{We use different labeling for some fields, but there is an obvious correspondence with Ref.~\cite{Barbier:2004ez}. For clarity we also denote the $SU(2)$ products with a dot.}
In this appendix, we recall some useful relations when writing the operators in terms of $SU(2)_L$ doublet fields, and then the relevant parts in the Lagrangian are presented.

Four-component Dirac spinors for leptons are constructed as
\begin{eqnarray}
	&& e = \left(
	\begin{array}{c}
		\psi_e\\
		\bar\psi_{e^c}
	\end{array}
	\right),~~~
	\nu = \left(
	\begin{array}{c}
		\psi_\nu\\
		\bar\psi_{\nu}
	\end{array}
	\right),
\end{eqnarray}
where the two-component Weyl spinor $\psi_i$ denotes the corresponding fermionic part in superfields.
Then, we may write the $SU(2)_L$ product for the doublet $L$ as
\begin{eqnarray}
	\bar L^c_i\cdot L_j  &=& \bar \nu^c_i P_L e_j - \bar e^c_iP_L \nu_j\,,
\end{eqnarray}
with $i$ and $j$ being the flavor indices.
We also note that in this notation $P_L e^c$ refers to the corresponding right-handed field, i.e., $\psi_{e^c}$.
We may define four component spinors for quarks in the same manner.

The MSSM superpotential is given by
\begin{eqnarray}
	W_{\rm MSSM} &=& \mu H_u \cdot H_d + y^e_{ij}H_d \cdot L_i E^c_j + y^d_{ij}H_d \cdot Q_i D^c_j + y^u_{ij} H_u \cdot Q_i U^c_j,
\end{eqnarray}
where $y^{e,u,d}$ are the Yukawa coupling matrices. The RPV trilinear couplings are given by
\begin{eqnarray}
	W^{(3)}_{\rm RPV} &=& \frac{1}{2}\lambda_{ijk}L_i \cdot L_j E^c_k + \lambda'_{ijk} L_i\cdot Q_j D^c_k + \frac{1}{2}\lambda''_{ijk}U^c_i D^c_j D^c_k,
\end{eqnarray}
and the corresponding Lagrangian in terms of four-component fermions becomes
\begin{eqnarray}
	{\cal L}_{L_iL_jE^c_k} &=& -\frac{1}{2}\lambda_{ijk}\left[\bar L^c_i\cdot L_j \tilde e^c_k+\bar e_k\tilde L_i\cdot L_j\right] + h.c.,\\
	{\cal L}_{L_iQ_jD^c_k} &=& - \lambda'_{ijk}
	\left[
		\tilde d^c_k \bar Q^c_j \cdot L_i+\bar d_k\tilde L_i\cdot Q_j+\bar d_k\tilde Q_j\cdot L_i
	\right] + h.c.,\\
	{\cal L}_{U^c_iD^c_jD^c_k} &=& -\frac{1}{2}\lambda''_{ijk}
	\left[
		\tilde u^c_{i} (\bar d_j P_L d^c_k) + \tilde d^c_{j} (\bar u_i P_L d^c_k) + \tilde d^c_{k} (\bar u_i P_L d^c_j)
	\right] + h.c.,
\end{eqnarray}
where $Q, L$ are quark, lepton doublets satisfying $\bar Q^c\cdot L =  \bar u^c P_L e-\bar d^c P_L \nu$, and $\lambda_{ijk}=-\lambda_{jik}$ and $\lambda''_{ijk}=-\lambda''_{jik}$ due to the gauge symmetry.

The soft supersymmetry breaking terms consist of the RPV part
\begin{eqnarray}
	-{\cal L}^{\rm soft}_{\rm RPV} = \frac{1}{2} A_{ijk} \tilde L_i \cdot \tilde L_j \tilde e^c_k + A'_{ijk} \tilde L_i \cdot \tilde Q_j \tilde d^c_k + \frac{1}{2} A''_{ijk} \tilde u^c_i \tilde d^c_j \tilde d^c_k + B_i h_u \cdot \tilde L_i + \widetilde m^2_{di}h^\dagger_d\tilde L_i + h.c.,
\end{eqnarray}
while the $R$-parity conserved part is given by
\begin{eqnarray}
	-{\cal L}^{\rm soft} &=& (m^2_{\tilde Q})_{ij} \tilde Q^\dagger_i \tilde Q_j + (m_{\tilde u^c}^2)_{ij}\tilde u^{c\dagger}_i\tilde u^c_j + (m_{\tilde d^c}^2)_{ij}\tilde d^{c\dagger}_i\tilde d^c_j + (m_{\tilde L}^2)_{ij}\tilde L^\dagger_i\tilde L_j + (m_{\tilde e^c}^2)_{ij}\tilde e^{c\dagger}_i\tilde e^c_j \nonumber\\
	&&
	+\left(
		A^e_{ij} h_d\cdot\tilde L_i\tilde e^c_j + A^d_{ij} h_d\cdot\tilde Q_i\tilde d^c_j - A^u_{ij} h_u\cdot\tilde Q_i\tilde u^c_j + h.c.
	\right)\nonumber\\
	&&
	+ \widetilde m_d^2 h^\dagger_d h_d + \widetilde m_u^2 h^\dagger_u h_u + (B h_u \cdot h_d + h.c.)\nonumber\\
	&&
	+ \frac{1}{2}M_1 \bar{\tilde B}\tilde B + \frac{1}{2}M_2\bar{\tilde W}^3\tilde W^3 + M_2 \bar{\tilde W}^+\tilde W^+ + \frac{1}{2} M_3 \bar{\tilde g}^a \tilde g^a,
\end{eqnarray}
where $\tilde Q$ and $\tilde L$ are squark and slepton doublets, respectively.

%%%%%%%%%%%%%%%%%%%%%%%%%%%%%%%%%%
%%%%%%%%%%% References %%%%%%%%%%%
%%%%%%%%%%%%%%%%%%%%%%%%%%%%%%%%%%

\end{document}